\documentclass[11pt]{article}

\usepackage[letterpaper,margin=1in]{geometry}
\usepackage{amsmath,amssymb,mathtools,bm}
\usepackage{microtype}
\usepackage{hyperref}
\usepackage[numbers,sort&compress]{natbib}
\usepackage{tikz}
\usetikzlibrary{positioning,calc,fit,arrows.meta}

\usepackage{amsfonts}

\title{\textbf{
A Stationary Composition Law for Schwinger-Keldysh Effective Actions}}
\author{Denis Comelli\\
\small INFN -- Sezione di Ferrara, Via Saragat 1, I-44122 Ferrara, Italy\\
\small \texttt{comelli@fe.infn.it}}
\date{}

\begin{document}

\maketitle
 
\begin{abstract}
 Let $W_{\cal O}[J]$ and $W_{\cal C}[J]$ denote the {\it connected generating
functionals} of an open   system and its closed system counterpart, and define 
 the {\it environment-induced} contribution $W_{\rm IF}[J]$ by the  difference:
$
    W_{\rm IF}[J]\equiv W_{\cal O}[J]-W_{\cal C}[J].
$
The corresponding Legendre transforms,
$\Gamma_{\cal O}$, $\Gamma_{\cal C}$, and $\Gamma_{\rm IF}$, do not obey an analogous additive relation
but satisfy a
 \emph{stationary composition law} given by: 
 \[ 
    \Gamma_{\cal O}[\Phi]
    =
    \operatorname*{Stat}_{\Psi}
    \left\{
        \Gamma_{\cal C}[\Psi]
        +
        \Gamma_{\rm IF}[\Phi-\Psi]
    \right\} 
\] where ${\rm Stat}$ denotes evaluation at a solution of the
stationarity condition with respect to $\Psi$.
This variational composition applies to both local and nonlocal Schwinger-Keldysh effective actions in nonequilibrium quantum field theory. We illustrate its linear and nonlinear realizations through quadratic theories and general 
time independent effective actions, respectively.

\end{abstract}
\tableofcontents

\section{Introduction}
 
 The Schwinger-Keldysh (SK), or in-in, formalism provides a real-time
framework for computing expectation values and correlation functions in
quantum systems prepared in a specified initial state. It originates in
the works of Schwinger and Keldysh
\cite{Schwinger:1960qe,Keldysh:1964ud} and was subsequently developed as
a general closed time path formulation of nonequilibrium quantum theory
\cite{Bakshi:1962dv,Chou:1984es,Jordan:1986ug}. Modern treatments may be
found in Refs.~\cite{Kamenev,Rammer:2007zz,Calzetta:2008iqa}. The
formalism is particularly suited to open quantum systems, where the
environmental degrees of freedom are integrated out to obtain a reduced
description of the system. This construction originates with the
Feynman--Vernon influence functional \cite{Feynman:1963fq} and leads to
an influence action encoding environment induced dissipation,
fluctuations, and nonlocal memory. Representative applications include
quantum Brownian motion and nonequilibrium field theory
\cite{Caldeira:1982iu,Hu:1991di,Paz:1990jg}, while a general account of
the open system framework is given in Ref.~\cite{Breuer:2007juk}.
Let $W_{\cal O}[J]$ and $W_{\cal C}[J]$ denote the 
{\it Connected Generating Functional} (CGF) of the open and corresponding closed systems, respectively.
We define the {\it environment-induced contribution} to the connected response
by
\begin{equation}
    W_{\rm IF}[J]
    \equiv
    W_{\cal O}[J]-W_{\cal C}[J],
    \label{eq:IntroductionConnectedDecomposition}
\end{equation}
Thus, $W_{\rm IF}$ is not, in general, the CGF of the isolated environment, but rather the change in the
system's CGF induced by its coupling to the
environment.
It is then natural to ask whether the corresponding effective actions
obey an analogous relation,
\begin{equation}
    \Gamma_{\cal O}[\Phi]
    \stackrel{?}{=}
    \Gamma_{\cal C}[\Phi]+\Gamma_{\rm IF}[\Phi].
    \label{eq:IntroductionNaiveSum}
\end{equation}
In general, Eq.~\eqref{eq:IntroductionNaiveSum} is not correct. The
obstruction follows directly from the Legendre transformation: each
            CGF defines its own background field,
\begin{equation}
    \Phi_{\cal O}[J]
    =
    \frac{\delta W_{\cal O}[J]}{\delta J},
    \qquad
    \Phi_{\cal C}[J]
    =
    \frac{\delta W_{\cal C}[J]}{\delta J},
    \qquad
    \Phi_{\rm IF}[J]
    =
    \frac{\delta W_{\rm IF}[J]}{\delta J}.
    \label{eq:IntroductionBackgrounds}
\end{equation}
Differentiating Eq.~\eqref{eq:IntroductionConnectedDecomposition} gives
\begin{equation}
    \Phi_{\cal O}[J]
    =
    \Phi_{\cal C}[J]+\Phi_{\rm IF}[J].
    \label{eq:IntroductionFieldDecomposition}
\end{equation}
Consequently, the additive relation between the CGFs implies an equality among their
Legendre transform values when each is evaluated at its own conjugate
background field:
\begin{equation}
    \Gamma_{\cal O}[\Phi_{\cal O}[J]]
    =
    \Gamma_{\cal C}[\Phi_{\cal C}[J]]
    +
    \Gamma_{\rm IF}[\Phi_{\rm IF}[J]].
    \label{eq:IntroductionDifferentBackgrounds}
\end{equation}
The problem is therefore not the additivity of the functional values,
but the mismatch between their natural arguments. In particular,
Eq.~\eqref{eq:IntroductionDifferentBackgrounds} does not provide an
expression for the open system effective action in terms of
$\Gamma_{\cal C}$ and $\Gamma_{\rm IF}$ evaluated at a prescribed common
background $\Phi$.
In this work, we shown that this mismatch is resolved by a stationary
composition. For a fixed total background $\Phi$, introduce an
auxiliary splitting field $\Psi$ and define
\begin{equation}
    \mathcal{F}[\Phi,\Psi]
    =
    \Gamma_{\cal C}[\Psi]
    +
    \Gamma_{\rm IF}[\Phi-\Psi].
    \label{eq:IntroductionAuxiliaryFunctional}
\end{equation}
The open system effective action is obtained as the {\it stationary value}
of this functional:
\begin{equation}
    \Gamma_{\cal O}[\Phi]
    =
   \operatorname*{Stat}_{\mathbf\Psi}
    \left\{
        \Gamma_{\cal C}[\Psi]
        +
        \Gamma_{\rm IF}[\Phi-\Psi]
    \right\}
    \equiv
    \left(
        \Gamma_{\cal C}\circledast\Gamma_{\rm IF}
    \right)[\Phi].
    \label{eq:IntroductionStationaryComposition}
\end{equation}
The auxiliary field partitions the total background into a
closed system contribution $\Psi$ and an environment induced
contribution $\Phi-\Psi$. Its {\it stationary value} is determined by solving  the stationarity equation
\begin{equation}
    \frac{\delta\Gamma_{\cal C}[\Psi]}{\delta\Psi}
    =
    \left.\frac{\delta\Gamma_{\rm IF}[\chi]}
         {\delta \chi}\right|_{\chi=\Phi-\Psi}\quad \to\quad \Psi=\Psi_{\star}[\Phi]
    \label{eq:IntroductionSourceMatching}
\end{equation}  giving the stationary $\Psi=\Psi_\star[\Phi]$ solution. Substitution into the auxiliary functional
then yields
\begin{equation}
    \Gamma_{\cal O}[\Phi]
    =
    \Gamma_{\cal C}[\Psi_\star[\Phi]]
    +
    \Gamma_{\rm IF}[\Phi-\Psi_\star[\Phi]].
\end{equation}
Through the inverse Legendre relations, this equation states that the
two contributions are generated by the same external source, (\ref{eq:source-matching}). The
stationary composition therefore replaces the explicit inversion of
the open system source field map (\ref{eq:phimap}) by a variational source-matching
condition.
Here the ${\rm Stat}$ operation 
does not imply minimization. 
In the Euclidean convex limit, the stationary
composition reduces to the usual {\it infimal convolution} 
associated with the Legendre-Fenchel transform \cite{Rock, Stromberg1994}.
We establish the principal algebraic properties of the stationary
composition, including commutativity and associativity. We also show
that the construction is covariant under different decompositions of
the connected response into closed system and environment induced
parts. Although the individual functionals $\Gamma_{\cal C}$ and
$\Gamma_{\rm IF}$ depend on this decomposition, their stationary
composition reconstructs the same open system effective action.
The formalism is illustrated through complementary linear and
nonlinear realizations. For quadratic integro-differential effective
actions, with kernels  $\mathbb A$ and $  \mathbb B$, stationary composition reduces to the parallel sum of the
corresponding response operators \cite{AndersonDuffin1969}, as shown in
Sec.~\ref{sec:realizations},
\begin{equation}
    \mathbb A\circledast  \mathbb B
    =
    \left( \mathbb A^{-1}+ \mathbb B^{-1}\right)^{-1}.
    \label{eq:IntroductionParallelSum}
\end{equation}
This result applies to local and nonlocal SK response operators,
including dissipative and memory kernels \cite{Breuer:2007juk}. For time independent
effective functionals, the stationary problem becomes a nonlinear
algebraic system expressing the balance between the generalized forces
generated by the closed system and influence contributions.
The paper is organized as follows. 
In Sec.~2 we review the SK
generating functionals for closed and open quantum systems.
 In Sec.~3
we introduce the corresponding effective actions, derive the stationary
composition law, and establish its commutativity, associativity and decomposition covariance. 
The quadratic and nonlinear
realizations of the construction are presented in Sec.~4, where also we discuss different strategies to compute the open system functionals. 
Section~5
contains our conclusions and outlines possible extensions.
   
    \section{SK Generating Functionals}

\subsection{Closed quantum systems}

For a system  prepared, at the initial time $t_i$, in the density matrix
$\hat{\rho}_i$, the expectation value of a time dependent operator
$\hat{O}(t)$ is defined as
\begin{equation}
\langle \hat{O}(t)\rangle
=
\mathrm{Tr}
\left[
\hat{\rho}_i\,
\hat U^\dagger(t,t_i)\,
\hat O(t_i)\,
\hat U(t,t_i)
\right],
\label{eq:expectation}
\end{equation}
where
\begin{equation}
\hat U(t,t_i)
=
{\cal T}
\exp
\left(
-i
\int_{t_i}^{t}
dt'\,
\hat H(t')
\right)
\end{equation}
is the time evolution operator and, $\hat H(t)$ is the Hamiltonian of the system and
$\mathcal T$ denotes chronological time ordering.
The simultaneous presence of the forward and backward evolution
operators requires a closed time contour extending from the initial
time $t_i$ to a final time $t_f$ and back again. The microscopic degrees of freedom
are consequently doubled, with independent field configurations
$q_1(t)$ and $q_2(t)$ living on the forward and backward branches,
respectively. Introducing independent  sources on the two branches,
$J_1(t)$ and $J_2(t)$, the SK action of a closed system
takes the form
\begin{equation}
S_{\cal C}[q_1,q_2;J_1,J_2]
=
S[q_1]
-
S[q_2]
+
\int dt\,
\Big(
J_1(t)\,q_1(t)
-
J_2(t)\,q_2(t)
\Big),
\label{eq:SKaction}
\end{equation}
where $S[q]$ denotes the {\it microscopic action} of the isolated system.
The corresponding generating functional is obtained by evolving the ket
of the initial density matrix along the forward branch, the bra along
the backward branch, and finally tracing over the common final state,
\begin{equation}
Z_{\cal C}[J_1,J_2]
=
\int dq_i\,dq_i'\;
\rho_i(q_i,q_i')
\int dq_f\;
{\cal A}[J_1;q_f,q_i]\;
{\cal A}^{*}[J_2;q_f,q_i'],
\label{eq:ZCdefinition}
\end{equation}
where
\begin{equation}
\rho_i(q_i,q_i')
=
\langle q_i|\hat\rho_i|q_i'\rangle
\end{equation}
are the matrix elements of the initial density matrix, while
\begin{equation}
{\cal A}[J;q_f,q_i]
=
\int_{q_i}^{q_f}
{\cal D}q\,
\exp\!\!
\left[
i\,S[q]
+
i\int dt\,J(t)\,q(t)
\right]
\label{eq:Amplitude}
\end{equation}
denotes the usual in-out transition amplitude in the presence of an
external source \cite{Kleinert:2004ev}.
Equation~(\ref{eq:ZCdefinition}) provides the microscopic starting point
for the  hierarchy of generating functionals constructed in the remainder of this work. In the following subsection we separate
the action into quadratic and interacting parts, allowing the generating
functional to be expressed as differential operators acting on a
Gaussian functional.
 
\subsection{Free and interacting generating functionals}
 
To construct perturbation theory, we separate the microscopic action
into a quadratic part and an interaction,
\begin{equation}
S[q] = S_0[q] + S_I[q],
\label{eq:split}
\end{equation}
where the free action $S_0$ is quadratic in the fields, so that the
corresponding path integral is Gaussian and can be evaluated exactly \cite{Kleinert:2004ev}.
The associated free transition amplitude is therefore
\begin{equation}
{\cal A}_0[J;q_f,q_i]
=
\int_{q_i}^{q_f}
{\cal D}q\,
\exp\!\!
\left[
i\,S_0[q]
+
i\int dt\,J(t)q(t)
\right].
\label{eq:A0}
\end{equation}
Using the standard functional-derivative representation,
$q(t)\rightarrow -i\,\delta/\delta J(t)$, immediately yields
\begin{equation}
{\cal A}[J;q_f,q_i]
=
e^{\,i\,S_I\!\left[-i\frac{\delta}{\delta J}\right]}\;
{\cal A}_0[J;q_f,q_i].
\label{eq:Aoperator}
\end{equation}
This naturally motivates the introduction of the free
SK generating functional
\begin{equation}
\begin{aligned}
Z_0[J_1,J_2]
=&
\int dq_i\,dq_i'\;
\rho_i(q_i,q_i')
\int dq_f\; {\cal A}_0[J_1;q_f,q_i]\,
{\cal A}_0^{*}[J_2;q_f,q_i'].
\end{aligned}
\label{eq:Z0def}
\end{equation}
Equivalently,
\begin{equation}
\begin{aligned}
Z_0[J_1,J_2]
=&
\int dq_i\,dq_i'\,
dq_f\; \rho_i(q_i,q_i')\,
\int_{q_i}^{q_f}{\cal D}q_1\,
\int_{q_i'}^{q_f}{\cal D}q_2
\\
&
\times
\exp
\Bigg[
i\,S_0[q_1]
-
i\,S_0[q_2]
+i\int_{t_i}^{t_f}dt
\left(
J_1\,q_1
-
J_2\,q_2
\right)
\Bigg].
\end{aligned}
\label{eq:Z0}
\end{equation}
The generating functional $Z_0$ contains the complete SK kinematics but no interactions. The only
approximation entering its definition is the replacement of the full
microscopic action by its quadratic part.
For later convenience, it is useful to exponentiate the initial density
matrix according to
\begin{equation}
\rho_i(q_i,q_i')
=
e^{\,i\,S_\rho[q_i, \,q_i']}  
\label{eq:Srho}
\end{equation}
which allows the boundary contribution to be incorporated into the
exponential together with the bulk action. In particular, for Gaussian
initial states, $S_\rho$ is itself quadratic and naturally combines with
the free action.
Substituting Eq.~(\ref{eq:Aoperator}) into the definition of the
SK generating functional finally gives
\begin{equation}
\boxed{
Z_{\cal C}[J_1,J_2]
=
e^{\,i\,\Delta S_I\!\left[-i\,\frac{\delta}{\delta_J}\right]}\;Z_0[J_1,J_2],}
\label{eq:masterclosed}
\end{equation}
where
\[
\Delta S_I[q_1,q_2]
=
S_I[q_1]-S_I[q_2].
\]
Equation~(\ref{eq:masterclosed}) is the SK analogue of the familiar interaction representation in ordinary quantum field theory: all interactions are encoded as differential operators acting on the Gaussian generating functional $Z_0$. It constitutes the fundamental representation of the
interacting closed system generating functional.
  
\subsection{Open quantum systems}
 
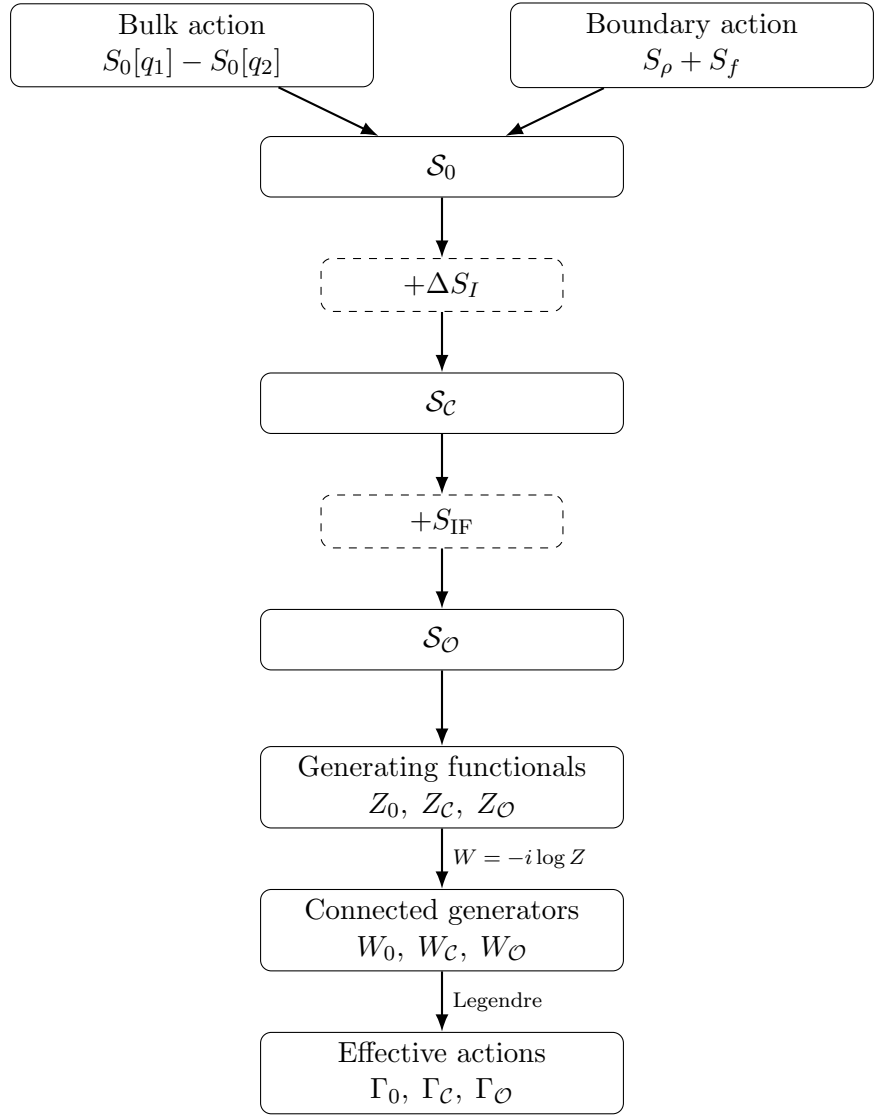
\begin{figure}[t]
\centering

\begin{tikzpicture}[
    >=Latex,
    node distance=8mm,
    every node/.style={align=center},
    box/.style={
        draw,
        rounded corners,
        minimum width=4.8cm,
        minimum height=8mm
    },
    action/.style={
        draw,
        dashed,
        rounded corners,
        minimum width=3.2cm,
        minimum height=7mm
    },
    arrow/.style={
        ->,
        thick
    }
]
 
\node[box] (bulk)
{Bulk action\\
$\displaystyle S_0[q_1]-S_0[q_2]$};

\node[box,right=1.8cm of bulk] (boundary)
{Boundary action\\
$\displaystyle S_\rho+S_f$};

\node[box,below=12mm of $(bulk)!0.5!(boundary)$] (S0)
{$\displaystyle \mathcal S_0$};

\node[action,below=8mm of S0] (DSI)
{$+\Delta S_I$};

\node[box,below=8mm of DSI] (SC)
{$\displaystyle \mathcal S_{\cal C}$};

\node[action,below=8mm of SC] (SIF)
{$+S_{\rm IF}$};

\node[box,below=8mm of SIF] (SO)
{$\displaystyle \mathcal S_{\cal O}$};

\node[box,below=10mm of SO] (Z)
{Generating functionals\\
$\displaystyle Z_0,\;Z_{\cal C},\;Z_{\cal O}$};

\node[box,below=8mm of Z] (W)
{Connected generators\\
$\displaystyle W_0,\;W_{\cal C},\;W_{\cal O}$};

\node[box,below=8mm of W] (G)
{Effective actions\\
$\displaystyle \Gamma_0,\;\Gamma_{\cal C},\;\Gamma_{\cal O}$};
 
\draw[arrow] (bulk)--(S0);
\draw[arrow] (boundary)--(S0);

\draw[arrow] (S0)--(DSI);
\draw[arrow] (DSI)--(SC);

\draw[arrow] (SC)--(SIF);
\draw[arrow] (SIF)--(SO);

\draw[arrow] (SO)--(Z);

\draw[arrow] (Z)--node[right]
{\scriptsize $W=-i\log Z$}(W);

\draw[arrow] (W)--node[right]
{\scriptsize Legendre}(G);

\end{tikzpicture}

\caption{
Hierarchical construction of the SK formalism.
The complete microscopic action consists of bulk and boundary
contributions. Adding the intrinsic interaction $\Delta S_I$
generates the closed system action, while the subsequent inclusion of
the influence action $S_{\rm IF}$ produces the open system action. The same
hierarchy propagates through the generating functionals, connected
generating functionals and effective actions.
}

\label{fig:Hierarchy}

\end{figure}

We now extend the previous construction to an open quantum system. At the
microscopic level, the system degrees of freedom $q$ are coupled to an
environment described by a set of variables $Q$. The total action can be
written schematically as
\begin{equation}
S_{\mathrm{tot}}[q,Q] = S[q] + S_E[Q] + S_{\mathrm{int}}[q,Q],
\label{eq:Stot}
\end{equation}
where $S[q]$ is the action of the isolated system, $S_E[Q]$ describes the
environment, and $S_{\mathrm{int}}$ their interaction.
Integrating out the environmental degrees of freedom produces the Feynman-Vernon influence action 
$S_{\mathrm{IF}}[q_1,q_2]$, which completely encodes the effect of the environment on the reduced SK dynamics of the system \cite{Feynman:1963fq,Caldeira:1982iu,Hu:1991di}. Although its explicit form depends on the microscopic model and on the initial state of the environment, the construction developed below is independent of these details.  
The SK action of the reduced open system therefore becomes
\begin{equation}
S_{\cal O}[q_1,q_2] = S[q_1] - S[q_2] + S_{\mathrm{IF}}[q_1,q_2]
=
S_{\cal C}[q_1,q_2]
+
S_{\mathrm{IF}}[q_1,q_2].
\label{eq:Sopen}
\end{equation}
Proceeding exactly as in the previous subsection, the influence action therefore acts as an additional interaction operator on the closed system generating functional.
\begin{equation}
\boxed{
Z_{\cal O}[J_1,J_2]
=
e^{\,i\,S_{\mathrm{IF}}
\!\left[
-i\frac{\delta}{\delta J_1},
\,i\frac{\delta}{\delta J_2}
\right]}\;
Z_{\cal C}[J_1,J_2].
}
\label{eq:Zopen}
\end{equation}
Combining Eq.~(\ref{eq:masterclosed}) with
Eq.~(\ref{eq:Zopen}) finally gives
\begin{equation}
\boxed{
Z_{\cal O}[J_1,J_2]
=
e^{\,i\,S_{\mathrm{IF}}
\!\left[
-i\frac{\delta}{\delta J_1},
\,i\frac{\delta}{\delta J_2}
\right]}\;
e^{\,i\,\Delta S_I
\!\left[
-i\frac{\delta}{\delta J_1},
\,i\frac{\delta}{\delta J_2}
\right]}\;
Z_0[J_1,J_2].
}
\label{eq:masteropen}
\end{equation}
Equations~(\ref{eq:masterclosed}) and
(\ref{eq:masteropen}) summarize the hierarchy of generating functionals, see our   Fig.~\ref{fig:Hierarchy},
\begin{equation}
Z_0
\;\longrightarrow\;
Z_{\cal C}
\;\longrightarrow\;
Z_{\cal O},\label{eq:Zhierar}
\end{equation}
The first step incorporates the intrinsic interactions of the isolated system, while the second accounts for the environment through the influence action. This hierarchical structure will be inherited by the CGF and, subsequently, by the effective actions.
The remainder of this work is devoted to understanding how this   structure evolves under successive Legendre transformations.

\subsection{Connected generating functionals}
 
The generating functionals introduced above contain the complete information about the quantum dynamics of the system. As in ordinary quantum field theory, it is convenient to work with the corresponding CGF,  defined by
\begin{equation}
W_a[J_1,J_2]
\equiv
-i\log Z_a[J_1,J_2],
\qquad
a=0,\,
{\cal C},\,
{\cal O},
\label{eq:defW}
\end{equation}
where the labels refer respectively to the quadratic theory ($a=0$), the
interacting closed system ($a={\cal C}$) and the open system ($a={\cal O}$).
 We therefore define the \emph{connected influence functional}  as the difference between the open and closed connected generators,
\begin{equation}
\boxed{
W_{\mathrm{IF}}[J_1,J_2]
\equiv
W_{\cal O}[J_1,J_2]
-
W_{\cal C}[J_1,J_2].
}
\label{eq:defWIF}
\end{equation}
Equivalently,
\begin{equation}
e^{\,i\,W_{\mathrm{IF} }[J_1,J_2]}
=
\frac{Z_{\cal O}[J_1,J_2]}{Z_{\cal C}[J_1,J_2]}=\frac{\int Dq\,e^{i\,S_{\rm IF}} \,e^{i\,S_{\cal C}+i\,Jq} }{\int Dq\, e^{i\,S_{\cal C}+i\,Jq} }
=
\left\langle
e^{\,i\,S_{\mathrm{IF}}}
\right\rangle_{{\cal C},J},
\label{eq:WIFaverage}
\end{equation}
where the average is performed with respect to the source-dependent
closed system path integral. Having established the main equations, we can now adopt the following compact notation
\begin{equation}
\mathbf{\Phi}
\equiv
\begin{pmatrix}
\Phi_1\\
\Phi_2
\end{pmatrix},
\qquad
\mathbf J
\equiv
\begin{pmatrix}
J_1\\
-J_2
\end{pmatrix},\qquad 
\frac{\delta}{\delta\mathbf J}
\equiv
\begin{pmatrix}\;
\;\frac{\delta}{\delta J_1}\\
-\frac{\delta}{\delta J_2}
\end{pmatrix}
\label{eq:SK-compact-notation}
\end{equation}
so that the source coupling is written as
\begin{equation}
\mathbf J\cdot\mathbf{\Phi}
= J_1\Phi_1-J_2\Phi_2.
\end{equation}
Accordingly, functionals operators $O$ of the two SK branches will be denoted compactly as
\begin{equation}
O[J_1,J_2]
\equiv
O[\mathbf J],
\qquad
O[\Phi_1,\Phi_2]
\equiv
O[\mathbf{\Phi}].
\end{equation}
We can use the shift identity
\begin{equation}
O\!\left[-i\frac{\delta}{\delta\mathbf J}\right]
Z_{\cal C}[\mathbf J]
=
Z_{\cal C}[\mathbf J]\;\;
O\!\left[
\mathbf\Phi_{\cal C}[\mathbf J]
-i\frac{\delta}{\delta\mathbf J}
\right]\cdot1.
\label{eq:shiftidentity}
\end{equation}
where
\[
\mathbf\Phi_{\cal C}[\mathbf J]
=
\frac{\delta W_{\cal C}[\mathbf J]}{\delta \mathbf J},
\]
so that Eq.~(\ref{eq:WIFaverage}) can be written entirely in terms of the
microscopic influence action as
\begin{equation}
\boxed{
W_{\mathrm{IF}}[\mathbf J]
=
-i
\log
\left[
e^{\,i\,S_{\mathrm{IF}}
\left[-i\,\frac{\delta}{\delta \mathbf J}+\mathbf\Phi_{\cal C}[\mathbf J]\right]}
\cdot1
\right].
}
\label{eq:WIFoperator}
\end{equation}
 While the microscopic influence action $S_{\rm IF}$ is obtained by integrating
out the environmental degrees of freedom and enters the reduced
microscopic SK action, Eqs.~\eqref{eq:WIFaverage}
and \eqref{eq:WIFoperator} shown that $W_{\rm IF}$ results from averaging
$e^{i\,S_{\rm IF}}$ over the source dependent closed system dynamics. It
therefore includes the interplay between intrinsic system fluctuations
and environmental backreaction and should be interpreted as the
environment-induced contribution to the connected generator of the
system, rather than as the CGF of the
isolated environment.
 The resulting additive decomposition of the open CGF is therefore
\begin{equation}
\boxed{
W_{\cal O}[\mathbf J]
=
W_{\cal C}[\mathbf J]
+
W_{\mathrm{IF}}[\mathbf J]
}\label{eq:defIF}
\end{equation}
and is the fundamental relation from which the
stationary composition of effective actions will be constructed in the
following sections.

\section{Effective Actions}
 
The CGF introduced in the previous section
provide a complete description of the real-time quantum dynamics through
their connected correlation functions. An equivalent and often more
convenient description is obtained through the Legendre transformation,
which defines the corresponding effective actions. Besides generating
the one-particle irreducible (1PI) vertex functions, the effective
actions determine the exact quantum equations of motion for the
macroscopic mean fields.
In the SK formalism, the Legendre transformation acts on
the complete generating functionals constructed from the microscopic
path integral. Consequently, the corresponding effective actions inherit
the full structure of the underlying SK theory. In
particular, the microscopic quantity entering the path integral is not
simply the bulk action of the system. It also contains the contribution
of the initial density matrix together with the final trace over the
common configuration of the forward and backward branches. It is
therefore useful to combine all these ingredients into a single
microscopic SK action before constructing the effective
actions.
  
\subsection{Microscopic SK actions}\label{subsec:microscopic-sk-actions}

The boundary conditions of the SK path integral are
introduced only once, namely in the free generating functional
$Z_0$, (\ref{eq:Z0}). They consist of the contribution of the initial density matrix (\ref{eq:Srho})
together with the final trace over the common configuration of the
forward and backward branches. The complete free microscopic action is
therefore \cite{BenTov:2021jsf,  Proko:2017,  Melo:2021mbd,  Barvinsky:2023jkl, Launay:2024trh}
\begin{equation}
\boxed{
\mathcal S_0
=
S_0[q_1]-S_0[q_2]
+
S_\rho
+
S_f,
}
\label{eq:S0complete}
\end{equation}
where $
S_\rho$ encodes the initial state~(\ref{eq:Srho}) and
\[
e^{i\,S_f}\equiv \lim_{\epsilon\to0^+}
\frac{1}{\sqrt{\pi\epsilon}}\,
\exp\!\left[
-\frac{\bigl(q_1(t_f)-q_2(t_f)\bigr)^2}{\epsilon}
\right]
=
\delta\!\bigl(q_1(t_f)-q_2(t_f)\bigr),
\]
implements the final trace \cite{Proko:2017}, \cite{Launay:2024trh}.
The interacting closed system and open system generating functionals are
constructed by successively adding the interaction and influence
actions,
\begin{align}
\mathcal S_{\cal C}
 =
\mathcal S_0
+
\Delta S_I,
\qquad
\mathcal S_{\cal O}
 =
\mathcal S_{\cal C}
+
S_{\rm IF}.\label{eq:Shier}
\end{align}
Thus the hierarchy
$
\mathcal S_0
\rightarrow
\mathcal S_{\cal C}
\rightarrow
\mathcal S_{\cal O}
$
mirrors the hierarchy (\ref{eq:Zhierar}).
The boundary conditions are fixed once through
$\mathcal S_0$ and are inherited by the subsequent generating
functionals and their corresponding effective actions.

\subsection{Legendre transformation}
 
For each of the CGF introduced in the
previous section,
 we define the corresponding effective action through the Legendre transformation
\begin{equation}
\Gamma_a[\mathbf \Phi]
=
W_a[\mathbf J ]
-
\int_{t_i}^{t_f}\!\!dt\;
\mathbf J(t)\cdot \mathbf \Phi_a (t),
\qquad
a=0,\,
{\cal C},\,
{\cal O},
\label{eq:Legendre}
\end{equation}
where the macroscopic mean fields are defined by
\begin{equation}
\mathbf \Phi_a (t)
=
\frac{\delta W_a[\mathbf J ]}{\delta \mathbf J (t)}.
\label{eq:MeanField}
\end{equation}
Taking the functional derivative of
Eq.~(\ref{eq:Legendre}) immediately gives the inverse Legendre
relations,
\begin{equation}
\frac{\delta\Gamma_a}{\delta\mathbf \Phi_a (t)}
=
-
\mathbf J (t),
\label{eq:InverseLegendre}
\end{equation}
which, in the absence of external sources, reduce to the stationary
conditions
\begin{equation}
\left.
\frac{\delta\Gamma_a}{\delta\mathbf \Phi_a (t)}
\right|_{\mathbf J =0}
\!\!\!\!=
0.
\label{eq:EOM}
\end{equation}
These are the exact quantum equations of motion for the corresponding
mean fields.
Since the CGF
$W_0$, $W_{\cal C}$ and $W_{\cal O}$ describe different microscopic
SK dynamics, their Legendre transformations are naturally
associated with different background field configurations (\ref{eq:MeanField}),
\begin{equation}
\mathbf\Phi_0[\mathbf J ]
\neq
\mathbf\Phi_{\cal C}[\mathbf J ]
\neq
\mathbf\Phi_{\cal O}[\mathbf J ],
\label{eq:DifferentFields}
\end{equation}
even when evaluated at identical external sources. Consequently, the
labels $0$, ${\cal C}$ and ${\cal O}$ will be retained throughout this
work in order to distinguish both the effective actions and their
associated background fields.
The existence of these different background fields has an important
consequence. While the CGFs admit the exact
decomposition
$
W_{\cal O}
=
W_{\cal C}
+
W_{\rm IF},
$
their corresponding effective actions are naturally evaluated at
different field configurations. This mismatch is precisely the
obstruction that prevents the additive decomposition of the connected
generating functionals from being transferred directly to the effective
actions. Resolving this mismatch constitutes the central problem
addressed in the remainder of this work.

\subsection{Exact functional representation and loop expansion of the effective action}
 
The construction of the effective action presented in this
subsection is completely independent of the particular microscopic
realization. We therefore omit the labels
$0$, ${\cal C}$ and ${\cal O}$  and denote by
$\mathcal{S}[q_1,q_2]$ the generic microscopic SK action. The resulting expressions apply equally to the
quadratic, closed and open systems discussed in this work.
For notational simplicity, all path integrals are understood to include
the integrations over the boundary configurations,
\[
\int dq_i\,dq_i'\,dq_f,
\]
together with the corresponding boundary conditions on the forward and
backward paths. Unless explicitly stated otherwise, these integrations
will not be written explicitly.
Starting from the definition of the Legendre transform,
Eq.~(\ref{eq:Legendre}), one obtains the exact functional equation satisfied by   the effective action 
\begin{equation}
e^{i\,\Gamma[\Phi_1,\Phi_2]}
=
\int
\mathcal Dq_1\,
\mathcal Dq_2\,
\exp
\left\{
i
\left[
\mathcal S[q_1,q_2]
+
\frac{\delta\Gamma}{\delta\Phi_1}
(\Phi_1-q_1)
-
\frac{\delta\Gamma}{\delta\Phi_2}
(\Phi_2-q_2)
\right]
\right\}.
\label{eq:ExactGamma}
\end{equation}
This is the SK counterpart of the standard DeWitt
representation of the effective action \cite{DeWitt:1964mxt} and provides
a common starting point for perturbative and non perturbative
approximations.
An important feature of Eq.~(\ref{eq:ExactGamma}) is that the microscopic
theory enters exclusively through the complete SK action
$\mathcal S$. Consequently, the same functional equation applies
without modification to the quadratic, closed and open systems. 
Although the functional equation
(\ref{eq:ExactGamma}) is exact, it cannot in general be solved
analytically. A systematic approximation is obtained by expanding the
path integral around the stationary configuration. To this end, the
microscopic fields $q_i$ are decomposed into a background configuration $\Phi_i(t)$ and
quantum fluctuations $\eta_i(t)$ as
\begin{equation}
q_i(t)
=
\Phi_i(t)
+
\eta_i(t),
\qquad
i=1,2,
\label{eq:FieldSplit}
\end{equation}
Expanding the complete microscopic SK action around the
background field gives
\begin{align}
\mathcal S[\mathbf \Phi+\mathbf \eta]
={} 
\mathcal S[\mathbf \Phi]
+
\int dt\,
\frac{\delta\mathcal S[\mathbf \Phi]}
{\delta\mathbf \Phi(t)}
\cdot \mathbf \eta(t)
+
\frac12
\int dt\,dt'\,
\mathbf \eta(t)\cdot
 {\cal A}(t,t';  \mathbf  \Phi)\cdot \mathbf 
\eta(t')
+
\mathcal O(\eta^3),
\label{eq:TaylorAction}
\end{align}
where repeated contour indices are summed and the quadratic fluctuation
kernel is
\begin{equation}
 {\cal A}(t,t';\mathbf \Phi)
=
\frac{\delta^2\mathcal S[\mathbf \Phi]}
{\delta\mathbf \Phi(t)\,
\delta\mathbf \Phi(t')}.
\label{eq:Kernel}
\end{equation}
In the exponent of Eq.~\eqref{eq:ExactGamma}, the term linear in the
fluctuation is cancelled order by order by the corresponding derivative
of the effective action. The remaining Gaussian integral gives the
one-loop contribution and the
effective action admits the formal loop expansion \cite{DeWitt:1964mxt, Jackiw:1974cv}
\begin{equation}
\Gamma[\mathbf \Phi]
=
\mathcal S[\mathbf \Phi]
+
\frac{i}{2}
\,\mathrm{Tr}
\log
\mathcal A[\mathbf \Phi]
+
\Gamma_{\ge2}[\mathbf \Phi],
\label{eq:LoopExpansion}
\end{equation}
where $\Gamma_{\ge2}$ denotes the sum of all contributions beginning at
two-loop order. These arise from the cubic and higher-order terms in the
expansion (\ref{eq:TaylorAction}) and contain the complete non-Gaussian
quantum corrections.
Equation~(\ref{eq:LoopExpansion}) has the same formal structure as the
conventional loop expansion of the effective action in quantum field
theory \cite{Kleinert:2004ev}. The essential difference in the SK formalism is
that the tree-level contribution is not simply the bare bulk action but
the complete microscopic SK action, including both the
bulk dynamics and the boundary contributions associated with the initial
density matrix and the final trace. Consequently, the loop expansion
describes quantum fluctuations around the complete real-time evolution
problem rather than around the bulk dynamics alone.
The general construction developed above applies directly to the three  microscopic actions $\mathcal S_0$,
$\mathcal S_{\cal C}$,
and
$\mathcal S_{\cal O}$ and 
  then the microscopic hierarchy (\ref{eq:Shier})
is   inherited by the corresponding effective actions,
\[
\Gamma_0
\longrightarrow
\Gamma_{\cal C}
\longrightarrow
\Gamma_{\cal O}.
\]
 Note that the label $a={\rm IF}$ is not included in this hierarchy.
The next subsection introduces the influence effective action. Together
with the effective actions constructed above, it provides the ingredients
required to derive the stationary composition of effective actions in the
following section.
  
\subsection{The influence effective action}
 
The additive decomposition of the CGF 
(\ref{eq:defIF})
naturally suggests introducing an effective action associated with the
connected influence functional. We therefore define the influence
effective action through the Legendre transformation
\begin{equation}
\boxed{
\Gamma_{\rm IF}[\mathbf{\Phi}_{\rm IF}]
=
W_{\rm IF}[\mathbf J]
-
\mathbf J\cdot
\mathbf{\Phi}_{\rm IF},
}
\label{eq:GammaIF}
\end{equation}
where
\begin{equation}
\mathbf{\Phi}_{\rm IF}[\mathbf J]
=
\frac{\delta W_{\rm IF}[\mathbf J]}
{\delta\mathbf J}.
\label{eq:PhiIF}
\end{equation}
The notation $\Gamma_{\rm IF}$ should not be interpreted as implying
that this functional is obtained directly from the microscopic
influence action $S_{\rm IF}$. The two objects belong to different
levels of the functional construction \cite{Su:1987pi}:
\begin{equation}
    S_{\rm IF}
    \longmapsto
    W_{\rm IF}[\mathbf J]
    =
    -i\log
    \left\langle
        e^{iS_{\rm IF}}
    \right\rangle_{{\cal C},\mathbf J}
    \longmapsto
    \Gamma_{\rm IF}[\mathbf\Phi_{\rm IF}].
\end{equation}
The first map averages the microscopic influence over the
source-dependent closed system dynamics, whereas the second is a
Legendre transformation. Therefore, $\Gamma_{\rm IF}$ describes the
influence-induced modification of the effective response of the
system and does not generally coincide with $S_{\rm IF}$. 
  For this reason, we refer to
$\Gamma_{\rm IF}$ more precisely as the \emph{influence-response
effective action}. In particular, even for a completely quadratic
theory, the kernel of $\Gamma_{\rm IF}$ does not coincide with the
kernel of the microscopic influence action $S_{\rm IF}$, see Sect.~(\ref{subsec:microscopic-quadratic-kernels}).
Differentiating Eq.~(\ref{eq:defIF}) immediately gives the
corresponding decomposition of the background fields,
\begin{equation}
\boxed{
\mathbf{\Phi}_{\cal O}[\mathbf J]
=
\mathbf{\Phi}_{\cal C}[\mathbf J]
+
\mathbf{\Phi}_{\rm IF}[\mathbf J],
}
\label{eq:PhiRelation}
\end{equation}
Here all three field profiles are evaluated at the same external source
$\mathbf J$, but they are generally different configurations.
The effective actions therefore satisfy
\begin{equation}
\Gamma_{\cal O}
\!\left[
\mathbf{\Phi}_{\cal O}[\mathbf J]
\right]
=
\Gamma_{\cal C}
\!\left[
\mathbf{\Phi}_{\cal C}[\mathbf J]
\right]
+
\Gamma_{\rm IF}
\!\left[
\mathbf{\Phi}_{\rm IF}[\mathbf J]
\right],
\label{eq:GammaDifferentFields}
\end{equation}
where each functional is evaluated at its own Legendre-conjugate
background field.
Equation~(\ref{eq:GammaDifferentFields}) is exact but should not be
interpreted as a decomposition of effective actions evaluated at a
common macroscopic configuration. Instead, it relates three effective
actions defined at different points in field space, all corresponding to
the same external source. This mismatch of background fields is the
fundamental obstacle to lifting the additive decomposition of the
CGF to the effective-action level.
For a prescribed open system background $\mathbf\Phi$, let
$\mathbf J_{\cal O}=\mathbf J_{\cal O}[\mathbf\Phi]$ denote its
Legendre-conjugate source, so that
\begin{equation}
    \mathbf\Phi
    \equiv
    \mathbf\Phi_{\cal O}[\mathbf J_{\cal O}].\label{eq:phimap}
\end{equation}
Evaluating Eq.~\eqref{eq:GammaDifferentFields} at
$\mathbf J=\mathbf J_{\cal O}$ and using
Eq.~\eqref{eq:PhiRelation}, one obtains
\begin{equation}
\boxed{
\Gamma_{\cal O}[\mathbf\Phi]
=
\Gamma_{\cal C}
\!\left[
    \mathbf\Phi_{\cal C}[\mathbf J_{\cal O}]
\right]
+
\Gamma_{\rm IF}
\!\left[
    \mathbf\Phi
    -
    \mathbf\Phi_{\cal C}[\mathbf J_{\cal O}]
\right].
}
\label{eq:GammaO-intermediate}
\end{equation}
This representation is exact, but its direct implementation requires
the inversion of the open system source field map in order to determine
$\mathbf J_{\cal O}[\mathbf\Phi]$ and hence
 $\mathbf\Phi_{\cal C}[\mathbf\Phi]=\mathbf\Phi_{\cal C}[\mathbf J_{\cal O}[\mathbf\Phi]]$. The stationary formulation
below replaces this explicit inversion by a source-matching condition.
  
\subsection{Stationary composition}
\label{subsec:stationary-composition}

For a prescribed total background $\mathbf\Phi$, introduce an
unconstrained splitting field $\mathbf\Psi$ and define
\begin{equation}
F[\mathbf\Phi,\mathbf\Psi]
\equiv
\Gamma_{\cal C}[\mathbf\Psi]
+
\Gamma_{\rm IF}[\mathbf\Phi-\mathbf\Psi].
\label{eq:F-functional}
\end{equation}
At fixed $\mathbf\Phi$, variation with respect to $\mathbf\Psi$ gives
\begin{equation}
\frac{\delta F[\mathbf\Phi,\mathbf\Psi]}
     {\delta\mathbf\Psi}
=
\frac{\delta\Gamma_{\cal C}[\mathbf\Psi]}
     {\delta\mathbf\Psi}
-
\left.
\frac{\delta\Gamma_{\rm IF}[\boldsymbol\chi]}
     {\delta\boldsymbol\chi}
\right|_{\boldsymbol\chi=\mathbf\Phi-\mathbf\Psi}.
\label{eq:F-variation}
\end{equation}
A stationary splitting
$\mathbf\Psi=\mathbf\Psi_\star[\mathbf\Phi]$ therefore satisfies
\begin{equation}
\left.
\frac{\delta\Gamma_{\cal C}[\mathbf\Psi]}
     {\delta\mathbf\Psi}
\right|_{\mathbf\Psi=\mathbf\Psi_\star}
=
\left.
\frac{\delta\Gamma_{\rm IF}[\boldsymbol\chi]}
     {\delta\boldsymbol\chi}
\right|_{\boldsymbol\chi=
\mathbf\Phi-\mathbf\Psi_\star}.
\label{eq:F-stationarity}
\end{equation}
Using the inverse Legendre relations,
\begin{equation}
\frac{\delta\Gamma_{\cal C}[\mathbf\Psi]}
     {\delta\mathbf\Psi}
=
-\mathbf J_{\cal C}[\mathbf\Psi],
\qquad
\frac{\delta\Gamma_{\rm IF}[\boldsymbol\chi]}
     {\delta\boldsymbol\chi}
=
-\mathbf J_{\rm IF}[\boldsymbol\chi],
\label{eq:inverse-Legendre-C-IF}
\end{equation}
the stationarity equation becomes
\begin{equation}
\mathbf J_{\cal C}[\mathbf\Psi_\star]
=
\mathbf J_{\rm IF}
[\mathbf\Phi-\mathbf\Psi_\star].
\label{eq:source-matching}
\end{equation}
Thus, the stationary decomposition is characterized by a common
Legendre-conjugate source for the closed system and influence-induced
parts of the total field.
To connect this stationary problem with the exact open system Legendre
transform, define
\begin{equation}
\mathbf\Psi_0[\mathbf\Phi]
\equiv
\mathbf\Phi_{\cal C}
\!\left[
\mathbf J_{\cal O}[\mathbf\Phi]
\right].
\label{eq:Psi-zero}
\end{equation}
Since
\begin{equation}
\mathbf\Phi
=
\mathbf\Phi_{\cal C}[\mathbf J_{\cal O}]
+
\mathbf\Phi_{\rm IF}[\mathbf J_{\cal O}],
\end{equation}
one has
\begin{equation}
\mathbf\Phi-\mathbf\Psi_0[\mathbf\Phi]
=
\mathbf\Phi_{\rm IF}
\!\left[
\mathbf J_{\cal O}[\mathbf\Phi]
\right].
\end{equation}
Assuming that the relevant Legendre maps are locally invertible, it
follows that
\begin{equation}
\mathbf J_{\cal C}[\mathbf\Psi_0]
=
\mathbf J_{\cal O}
=
\mathbf J_{\rm IF}
[\mathbf\Phi-\mathbf\Psi_0].
\end{equation}
Consequently, $\mathbf\Psi_0$ satisfies the source-matching condition
\eqref{eq:source-matching} and is a stationary configuration of
$F[\mathbf\Phi,\mathbf\Psi]$.
On the stationary branch determined by the local open system Legendre
transform, one therefore has
\begin{equation}
\mathbf\Psi_\star[\mathbf\Phi]
=
\mathbf\Phi_{\cal C}
\!\left[
\mathbf J_{\cal O}[\mathbf\Phi]
\right],
\qquad
\mathbf\Phi-\mathbf\Psi_\star[\mathbf\Phi]
=
\mathbf\Phi_{\rm IF}
\!\left[
\mathbf J_{\cal O}[\mathbf\Phi]
\right].
\label{eq:stationary-identification}
\end{equation}
Substitution into Eq.~\eqref{eq:F-functional} reproduces
Eq.~\eqref{eq:GammaO-intermediate}. Hence, the open system effective
action is given by
\begin{equation}
\boxed{
\Gamma_{\cal O}[\mathbf\Phi]
=
\operatorname*{Stat}_{\mathbf\Psi}
\left\{
\Gamma_{\cal C}[\mathbf\Psi]
+
\Gamma_{\rm IF}[\mathbf\Phi-\mathbf\Psi]
\right\}.
}
\label{eq:GammaO-stationary}
\end{equation}
Here, $\operatorname*{Stat}_{\mathbf\Psi}$ denotes evaluation at a
solution of the stationarity condition with respect to $\mathbf\Psi$;
it does not imply minimization.
This result motivates the definition of the stationary composition of
two effective actions:
\begin{equation}
\left(
\Gamma_1\circledast\Gamma_2
\right)[\mathbf\Phi]
\equiv
\operatorname*{Stat}_{\mathbf\Psi}
\left\{
\Gamma_1[\mathbf\Psi]
+
\Gamma_2[\mathbf\Phi-\mathbf\Psi]
\right\}.
\label{eq:stationary-composition-definition}
\end{equation}
The central composition law then takes the compact form
\begin{equation}
\boxed{
\Gamma_{\cal O}
=
\Gamma_{\cal C}\circledast\Gamma_{\rm IF}.
}
\label{eq:central-composition}
\end{equation}
The auxiliary field $\mathbf\Psi_\star$ represents the closed system
contribution to the total response, whereas
$\mathbf\Phi-\mathbf\Psi_\star$ represents the corresponding
environment-induced contribution. These are not independent physical
fields, but the two components of the total background selected by the
common source condition \eqref{eq:source-matching}.
The Hessian of the auxiliary functional with respect to the splitting
field is
\begin{align}
\mathcal H_\star[\mathbf\Phi]
&\equiv
\left.
\frac{\delta^2F[\mathbf\Phi,\mathbf\Psi]}
     {\delta\mathbf\Psi\,\delta\mathbf\Psi}
\right|_{\mathbf\Psi=\mathbf\Psi_\star}
\!\!\!=
\left.
\frac{\delta^2\Gamma_{\cal C}[\mathbf\Psi]}
     {\delta\mathbf\Psi\,\delta\mathbf\Psi}
\right|_{\mathbf\Psi=\mathbf\Psi_\star}
+
\left.
\frac{\delta^2\Gamma_{\rm IF}[\boldsymbol\chi]}
     {\delta\boldsymbol\chi\,\delta\boldsymbol\chi}
\right|_{\boldsymbol\chi=
\mathbf\Phi-\mathbf\Psi_\star}.
\label{eq:F-nondegeneracy}
\end{align}
Under the standard local regularity assumptions, invertibility of
$\mathcal H_\star$ ensures that the implicit-function theorem determines
a locally unique stationary branch
$\mathbf\Psi_\star=\mathbf\Psi_\star[\mathbf\Phi]$.
If the stationarity condition admits additional solutions, stationarity
alone does not select among them. The solution identified in
Eq.~\eqref{eq:stationary-identification} is guaranteed to reproduce the
open system effective action. Other stationary branches require an
additional selection prescription, and their analysis lies beyond the
scope of the present work.
The stationary composition law may be viewed as the SK counterpart of
a standard result of convex duality. In the Euclidean convex setting,
under suitable regularity assumptions, the Legendre--Fenchel transform
maps the sum of two functionals to the infimal convolution of their
duals \cite{Rock,Stromberg1994}. In the SK setting, however, the
effective actions are generally complex and need not be convex, so the
infimum is replaced by stationary evaluation. The operation
$\circledast$ may therefore be regarded as a formal SK analogue of
infimal convolution. In the Euclidean convex limit, whenever the
infimum is attained, stationary evaluation reduces to minimization and
the two operations coincide.

       \subsubsection{Algebraic properties}
The stationary composition inherits two main algebraic
properties from its variational definition.

\begin{itemize}
\item\textbf{ Commutativity }:
The stationary composition is commutative,
\begin{equation}
\boxed{
\Gamma_1\circledast\Gamma_2
=
\Gamma_2\circledast\Gamma_1.
}
\label{eq:commutativity}
\end{equation}

\noindent
\emph{Proof.}
Starting from the definition,
\[
(\Gamma_1\circledast\Gamma_2)[\mathbf \Phi]
=
\operatorname*{Stat}_{\mathbf\Psi}
\left\{
\Gamma_1[\mathbf \Psi]
+
\Gamma_2[\mathbf \Phi-\mathbf \Psi]
\right\},
\]
introduce the new variable
$
\mathbf \Xi=\mathbf \Phi-\mathbf \Psi$.
Since this transformation is invertible, stationary points are preserved,
and
\[
\Gamma_1[\mathbf \Psi]
+
\Gamma_2[\mathbf \Phi-\mathbf \Psi]
=
\Gamma_2[\mathbf \Xi]
+
\Gamma_1[\mathbf \Phi-\mathbf \Xi].
\]
Hence,
\[
(\Gamma_1\circledast\Gamma_2)[\mathbf \Phi]
=
(\Gamma_2\circledast\Gamma_1)[\mathbf \Phi],
\]
which proves Eq.~\eqref{eq:commutativity}.
\hfill$\square$

\item 
\textbf{ Associativity}:
The stationary composition is associative,
\begin{equation}
\boxed{
(\Gamma_1\circledast\Gamma_2)\circledast\Gamma_3
=
\Gamma_1\circledast
(\Gamma_2\circledast\Gamma_3).
}
\label{eq:associativity}
\end{equation}

\noindent
\emph{Proof.}
Applying the definition twice gives
\[
\bigl((\Gamma_1\circledast\Gamma_2)\circledast\Gamma_3\bigr)[\mathbf \Phi]
=\operatorname*{Stat}_{\boldsymbol{\chi},\mathbf \Psi}
\left\{
\Gamma_1[\mathbf \Psi]
+
\Gamma_2[\boldsymbol{\chi}-\mathbf \Psi]
+
\Gamma_3[\mathbf \Phi-\boldsymbol{\chi}]
\right\}.
\]
Introducing the variable
$
\mathbf \Xi=\boldsymbol{\chi}-\mathbf \Psi$,
one obtains
\[ \operatorname*{Stat}_{\mathbf \Xi,\mathbf \Psi}
\left\{
\Gamma_1[\mathbf \Psi]
+
\Gamma_2[\mathbf \Xi]
+
\Gamma_3[\mathbf \Phi-\mathbf \Psi-\mathbf \Xi]
\right\},
\]
which is  the stationary representation obtained from
\(
\Gamma_1\circledast(\Gamma_2\circledast\Gamma_3)
\).
Therefore,
Eq.~\eqref{eq:associativity} follows.
Associativity has a natural interpretation when the connected
generating functional contains several environment-induced
contributions. In this setting, $\Gamma_2$ and $\Gamma_3$ are not
effective actions of the baths themselves, nor are they additive
contributions to $\Gamma_{\cal O}$. They are the Legendre transforms of
the corresponding bath induced connected functionals. Their stationary
composition
\[
    \Gamma_{23}
    \equiv
    \Gamma_2\circledast\Gamma_3
\]
represents their combined contribution in the Legendre-dual
description. 
Associativity
implies that this combined contribution may be composed with the
closed system effective action independently of the order of composition provided the corresponding stationary branches are compatible.
\item 
\textbf{ Composition of multiple contributions}:
Suppose that the open system connected generator admits the additive
decomposition
\begin{equation}
    W_{\cal O}[\mathbf J]
    =
    W_{\cal C}[\mathbf J]
    +
    \sum_{i=1}^{n}W_i[\mathbf J].
    \label{eq:multiple-W-decomposition}
\end{equation}
 The additive decomposition in
Eq.~\eqref{eq:multiple-W-decomposition} is an assumption at the level
of connected generators. It does not follow in general from an
additive microscopic influence action, since the closed system average
can generate mixed connected cumulants of the different influence
sectors.
Let $\Gamma_i$ denote the Legendre transform of $W_i$. Repeated
application of the stationary composition gives
\begin{equation}
\boxed{
    \Gamma_{\cal O}[\mathbf\Phi]
    =\operatorname*{Stat}_{ \mathbf\Psi_{\cal C}+\sum_{i=1}^{n}\mathbf\Psi_i  =\mathbf\Phi} 
    \left\{
      \Gamma_{\cal C}[\mathbf\Psi_{\cal C}]
      +
      \sum_{i=1}^{n}\Gamma_i[\mathbf\Psi_i]
    \right\}.
}
\label{eq:multiple-stationary-composition}
\end{equation}
Equivalently, eliminating the closed system contribution gives
\begin{equation}
    \Gamma_{\cal O}[\mathbf\Phi]
    =\operatorname*{Stat}_{\mathbf\Psi_1,\ldots,\mathbf\Psi_n}
    \left\{
      \Gamma_{\cal C}
      \left[
        \mathbf\Phi-\sum_{i=1}^{n}\mathbf\Psi_i
      \right]
      +
      \sum_{i=1}^{n}\Gamma_i[\mathbf\Psi_i]
    \right\}.
\label{eq:multiple-stationary-composition-reduced}
\end{equation}
The stationary conditions match the Legendre-conjugate sources of all
contributions,
\begin{equation}
    \mathbf J_{\cal C}[\mathbf\Psi_{\cal C}]
    =
    \mathbf J_1[\mathbf\Psi_1]
    =
    \cdots
    =
    \mathbf J_n[\mathbf\Psi_n].
\label{eq:multiple-source-matching}
\end{equation}
Associativity ensures that the same open system effective action is
obtained independently of the order in which the individual
contributions are composed.
\end{itemize}
The stationary composition therefore defines a commutative and
associative binary operation on the class of effective actions for which
the corresponding stationary branches exist.

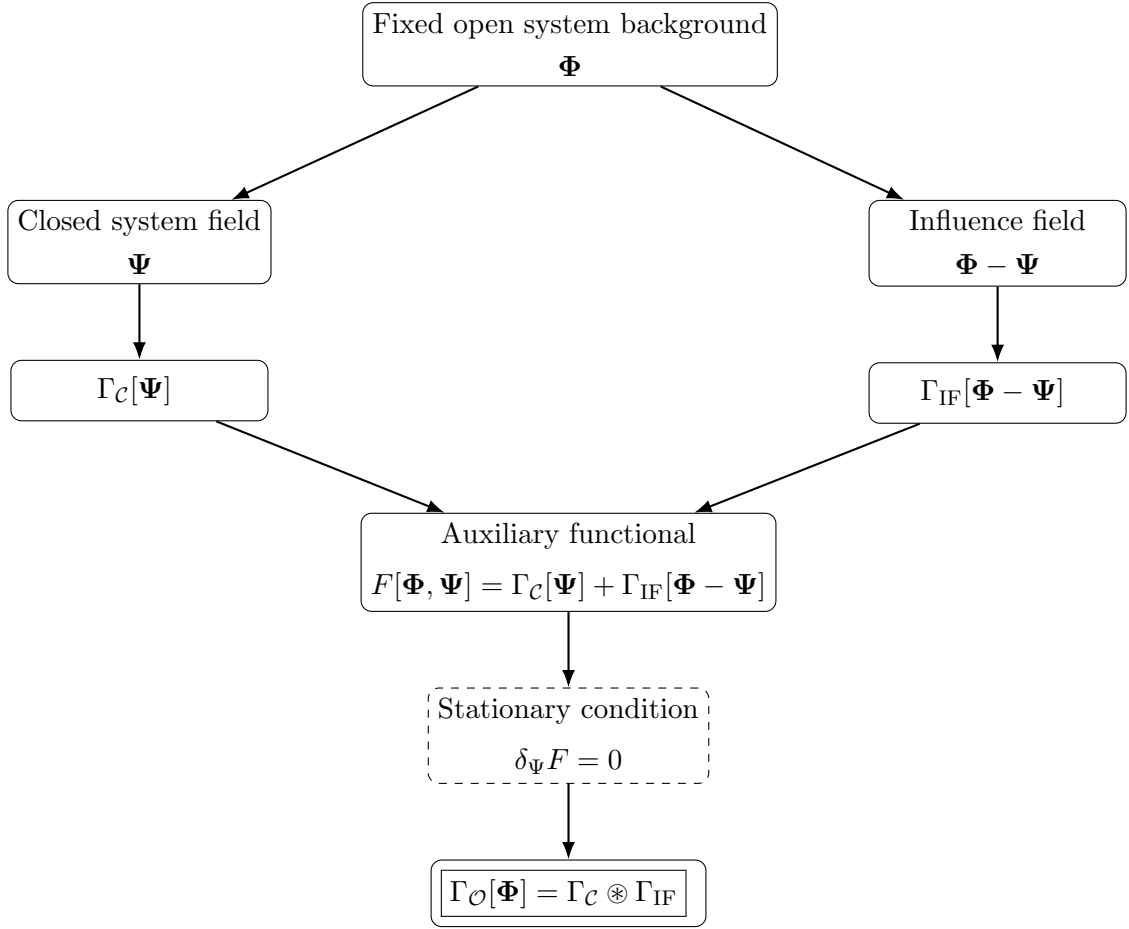
\begin{figure}[t]
\centering

\begin{tikzpicture}[
    >=Latex,
    node distance=10mm,
    every node/.style={align=center},
    field/.style={
        draw,
        rounded corners,
        minimum width=3.4cm,
        minimum height=8mm
    },
    action/.style={
        draw,
        dashed,
        rounded corners,
        minimum width=2.8cm,
        minimum height=8mm
    },
    arrow/.style={
        ->,
        thick
    }
]
 
\node[field] (Phi)
{
Fixed open system background
\\[1mm]
$\mathbf{\Phi}$
};

\node[field,
below left=15mm and 12mm of Phi] (Psi)
{
Closed system field
\\[1mm]
$\mathbf{\Psi}$
};

\node[field,
below right=15mm and 12mm of Phi] (Chi)
{
Influence field
\\[1mm]
$\mathbf{\Phi}-\mathbf{\Psi}$
};

\node[field,
below=10mm of Psi] (GammaC)
{
$\Gamma_{\cal C}[\mathbf{\Psi}]$
};

\node[field,
below=10mm of Chi] (GammaIF)
{
$\Gamma_{\rm IF}[\mathbf{\Phi}-\mathbf{\Psi}]$
};

\node[field,
below=16mm of $(GammaC)!0.5!(GammaIF)$] (F)
{
Auxiliary functional
\\[2mm]
$\displaystyle
F[\mathbf{\Phi},\mathbf{\Psi}]
=
\Gamma_{\cal C}[\mathbf{\Psi}]
+
\Gamma_{\rm IF}[\mathbf{\Phi}-\mathbf{\Psi}]
$
};

\node[action,
below=10mm of F] (Stat)
{
Stationary condition
\\[2mm]
$\displaystyle
\delta_\Psi F=0
$
};

\node[field,
below=10mm of Stat] (GammaO)
{
$\boxed{
\Gamma_{\cal O}[\mathbf{\Phi}]
=
\Gamma_{\cal C}
\circledast
\Gamma_{\rm IF}
}$
};

\draw[arrow] (Phi)--(Psi);
\draw[arrow] (Phi)--(Chi);

\draw[arrow] (Psi)--(GammaC);
\draw[arrow] (Chi)--(GammaIF);

\draw[arrow] (GammaC)--(F);
\draw[arrow] (GammaIF)--(F);

\draw[arrow] (F)--(Stat);

\draw[arrow] (Stat)--(GammaO);

\end{tikzpicture}

\caption{
Stationary composition of effective actions.
The total macroscopic background field $\mathbf{\Phi}$ is decomposed
into a closed system contribution $\mathbf{\Psi}$ and an environmental
contribution $\mathbf{\Phi}-\mathbf{\Psi}$. Their corresponding
effective actions define the auxiliary functional
$F[\mathbf{\Phi},\mathbf{\Psi}]$. The stationary decomposition is selected
by the stationary condition
$\delta_{\mathbf{\Psi}}F=0$, yielding the stationary composition
$\Gamma_{\cal O}
=
\Gamma_{\cal C}\circledast\Gamma_{\rm IF}$.
}

\label{fig:StationaryComposition}

\end{figure}

\subsubsection{Decomposition covariance}
\label{sec:decomposition-covariance}
 
The decomposition of the open system CGF (\ref{eq:defIF})
is not unique. Indeed, for an arbitrary source functional
$X[\mathbf J]$ (note that we restrict to transformations   for which the relevant Legendre
maps remain locally invertible and the corresponding stationary branches
exist),  consider the transformations
\begin{equation}
W_{\cal C}[\mathbf J]
\longrightarrow
W_{\cal C}^{\,X}[\mathbf J]
=
W_{\cal C}[\mathbf J]
+
X[\mathbf J],\quad
W_{\rm IF}[\mathbf J]
\longrightarrow
W_{\rm IF}^{\,X}[\mathbf J]
=
W_{\rm IF}[\mathbf J]
-
X[\mathbf J].
\label{eq:WIF-shift}
\end{equation}
Their sum is unchanged,
\begin{equation}
W_{\cal O}^{\,X}[\mathbf J]
=
W_{\cal C}^{\,X}[\mathbf J]
+
W_{\rm IF}^{\,X}[\mathbf J]
=
W_{\cal O}[\mathbf J].
\label{eq:WO-shift-invariance}
\end{equation}
We refer to the freedom
(\ref{eq:WIF-shift}) as
\emph{decomposition covariance}. It expresses the fact that the
separation of the connected response into intrinsic and
environment-induced contributions depends on the chosen decomposition,
whereas their sum does not.
The corresponding background fields transform according to
\begin{align}
\mathbf{\Phi}_{\cal C}^{\,X}[\mathbf J]
&=
\mathbf{\Phi}_{\cal C}[\mathbf J]
+
\frac{\delta X[\mathbf J]}
{\delta\mathbf J},
\qquad
\mathbf{\Phi}_{\rm IF}^{\,X}[\mathbf J]
=
\mathbf{\Phi}_{\rm IF}[\mathbf J]
-
\frac{\delta X[\mathbf J]}
{\delta\mathbf J},
\end{align}
so that the total open system background field remains invariant,
\begin{equation}
\mathbf{\Phi}_{\cal O}^{\,X}[\mathbf J]
=
\mathbf{\Phi}_{\cal C}^{\,X}[\mathbf J]
+
\mathbf{\Phi}_{\rm IF}^{\,X}[\mathbf J]
=
\mathbf{\Phi}_{\cal O}[\mathbf J].
\label{eq:PhiO-shift-invariance}
\end{equation}
Let
$\Gamma_{\cal C}^{\,X}$ and
$\Gamma_{\rm IF}^{\,X}$
denote the Legendre transforms of
$W_{\cal C}^{\,X}$ and
$W_{\rm IF}^{\,X}$,
respectively. Since their connected generators still satisfy Eq.(\ref{eq:WO-shift-invariance})
the stationary composition derived in the previous section immediately
gives
\begin{equation}
\Gamma_{\cal O}
=
\Gamma_{\cal C}^{\,X}
\circledast
\Gamma_{\rm IF}^{\,X}
=
\Gamma_{\cal C}
\circledast
\Gamma_{\rm IF}.
\label{eq:Gamma-decomposition-covariance}
\end{equation}
 The individual effective actions $\Gamma_{\cal C}$ and
$\Gamma_{\rm IF}$ depend on the chosen decomposition, whereas their
stationary composition, $\Gamma_{\cal O}$, is invariant. 
Consequently,
all quantities derived from the complete effective action or,
equivalently, from $W_{\cal O}$ are invariant under the transformations
in Eq.~\eqref{eq:WIF-shift}. These include the open system mean field,
the connected correlation functions, and the corresponding
one-particle-irreducible vertex functions.
This decomposition covariance is a property of the reduced effective
description, not of the underlying microscopic theory. A specified
microscopic model naturally determines a decomposition into closed system
and environment-induced contributions. Once only the reduced generating
functional $W_{\cal O}$ is retained, however, this decomposition is no
longer unique: any pair $(W_{\cal C},W_{\rm IF})$ satisfying
$W_{\cal O}=W_{\cal C}+W_{\rm IF}$ yields the same $\Gamma_{\cal O}$
through stationary composition.

 \section{Linear and nonlinear realizations}
\label{sec:realizations}

\subsection{Quadratic integro-differential effective actions}

Quadratic effective actions provide the simplest nontrivial realization
of the stationary composition principle. Although the corresponding
stationary equation is linear and therefore admits an explicit solution,
the quadratic approximation already encompasses a broad class of physical
systems, including local effective field theories, non-Markovian memory
effects \cite{Breuer:2007juk}, dissipative open system dynamics \cite{Calzetta:2008iqa} and, more generally, linear
response theory \cite{Kubo:1957mj}. 
We consider the quadratic effective actions
\begin{align}
\Gamma_{\cal C}[\mathbf\Psi]
&=
\frac12
\mathbf\Psi\cdot  \mathbb A\cdot\mathbf\Psi,
\qquad
\Gamma_{\rm IF}[ \mathbf\Psi]
=
\frac12\,
  \mathbf\Psi \cdot \mathbb B\cdot  \mathbf\Psi ,\label{eq:QUADGAMMA}
\end{align}
where the $2\times 2$ matrices $\mathbb A$ and $\mathbb B$ are   the response operators associated with the
closed system and influence effective actions, respectively. 
Depending
on the physical application, these operators may be local differential
operators, nonlocal memory kernels, or, after a Keldysh rotation,
matrices of retarded, advanced, and Keldysh components.
 The derivation below depends only on
their linearity and is therefore independent of their particular
realization. In particular the operator $\mathbb B$,
  should not be identified with the kernel of the microscopic
influence action as shown 
in Sec.~\ref{subsec:microscopic-quadratic-kernels}. 
According to the stationary composition principle, the open system
effective action is obtained from the auxiliary functional
\begin{equation}
F[\mathbf\Phi,\mathbf\Psi]
=
\Gamma_{\cal C}[\mathbf\Psi]
+
\Gamma_{\rm IF}[\mathbf\Phi-\mathbf\Psi] =
\frac12
\mathbf\Psi\cdot \mathbb A\cdot\mathbf\Psi
+
\frac12
(\mathbf\Phi-\mathbf\Psi)\cdot \mathbb B\cdot(\mathbf\Phi-\mathbf\Psi).
\end{equation}
The stationary configuration is given by the solution of the following Eq.
\begin{equation}
(\mathbb A+\mathbb B)\cdot \mathbf\Psi
=
\mathbb B\cdot \mathbf\Phi
\label{eq:linear_stationary}
\end{equation}
 that requires the inversion of the auxiliary Hessian   
\begin{equation}
F_{ \mathbf\Psi\mathbf\Psi}
=
\mathbb A+\mathbb B.
\end{equation}
Its invertibility guarantees the existence of a locally unique
stationary splitting given by
\begin{equation}
{\mathbf\Psi}_\star
=
(\mathbb A+\mathbb B)^{-1}\cdot \mathbb B\cdot \mathbf\Phi.
\label{eq:quadratic_solution}
\end{equation}
Substituting Eq.~(\ref{eq:quadratic_solution}) into the auxiliary
functional yields the composed effective action
\begin{equation}
\Gamma_{\cal O}[\mathbf\Phi]
=
\frac12\,
\mathbf\Phi\cdot(\mathbb A^{-1}+\mathbb B^{-1})^{-1}
\cdot\mathbf\Phi,\label{eq:quadratic_GammaO}
\end{equation}
so that we can translate the $\circledast$ operation to quadratic operators as
\begin{equation}
\Gamma_{\cal C} \circledast \Gamma_{\rm IF}=\Gamma_{\cal O}
\quad \to\quad
    \mathbb A\circledast  \mathbb B
    =
    \left( \mathbb A^{-1}+ \mathbb B^{-1}\right)^{-1}.
    \label{eq:IntroductionParallelSum1}
\end{equation}
The quadratic example provides the general linear realization of the
stationary composition principle.

    \subsection{Microscopic interpretation of the quadratic kernels}
\label{subsec:microscopic-quadratic-kernels}
The quadratic realization discussed above assumes that the effective
actions $\Gamma_{\cal C}$ and $\Gamma_{\rm IF}$ are known. We now
consider the complementary perspective in which the microscopic
closed system ${\cal S}_{\cal C}$ and influence actions $S_{\rm IF}$ are instead taken as the starting
point. This provides a direct relation between the microscopic
influence kernel and the kernel entering the influence effective action.
Consider the complete quadratic microscopic SK actions
\begin{equation}
    \mathcal S_{\cal C}[\mathbf q]
    =
    \frac{1}{2}\,
    \mathbf q\cdot \mathbb K_{\cal C}\cdot\mathbf q,
    \qquad
    S_{\rm IF}[\mathbf q]
    =
    \frac{1}{2}\,
    \mathbf q\cdot \mathbb   K_{\rm IF}\cdot\mathbf q.
    \label{eq:microscopic-quadratic-actions}
\end{equation} parametrized by the $2\times 2$ matrix kernels $\mathbb K_{\cal C}$ and $\mathbb K_{\rm IF}$.
The complete open system action is therefore
\begin{equation}
    \mathcal S_{\cal O}[\mathbf q]
    =
    \frac{1}{2}\,
    \mathbf q\cdot
    \left(\mathbb K_{\cal C}+\mathbb K_{\rm IF}\right)
    \cdot\mathbf q.
    \label{eq:microscopic-open quadratic-action}
\end{equation}
For a quadratic theory, the loop expansion terminates at the classical
level (\ref{eq:LoopExpansion}), up to field-independent contributions. Consequently,
\begin{equation}
    \Gamma_{\cal C}[\Phi]
    =
    \frac{1}{2}\,\Phi\cdot \mathbb K_{\cal C}\cdot\Phi,
    \qquad
    \Gamma_{\cal O}[\Phi]
    =
    \frac{1}{2}\,\Phi\cdot
    \left(\mathbb K_{\cal C}+\mathbb K_{\rm IF}\right)
    \cdot\Phi.
    \label{eq:quadratic-classical-effective-actions}
\end{equation}
Thus, in the notation of the previous quadratic realization (\ref{eq:QUADGAMMA}), we get
\begin{equation}
    \mathbb A=\mathbb K_{\cal C}.
    \label{eq:A-microscopic-kernel}
\end{equation} and using the stationary composition of Eq.~(\ref{eq:quadratic_GammaO})
  compared with
Eq.~\eqref{eq:quadratic-classical-effective-actions} it requires
\begin{equation}
    \mathbb B^{-1}
    =
    \left(\mathbb K_{\cal C}+\mathbb K_{\rm IF}\right)^{-1}
    -
    \mathbb K_{\cal C}^{-1}.
\label{eq:B-from-microscopic-kernels}
\end{equation}
This relation holds whenever the corresponding operators and their
differences admit the required inverses.
Equation~\eqref{eq:B-from-microscopic-kernels} shows that the kernel
$\mathbb B$ of $\Gamma_{\rm IF}$ does not coincide with the microscopic
influence kernel $\mathbb K_{\rm IF}$. 
This result explicitly confirms that the microscopic influence kernel
and the kernel of the influence-response effective action are generally
distinct.

\paragraph{Partially quadratic cases.}

The relation between the microscopic influence action and the connected
influence functional simplifies also when either the closed system action or
the influence action is quadratic. 
\begin{itemize}
\item If the complete closed system SK action  ${\cal S}_{\cal C}$ is quadratic 
(\ref{eq:microscopic-quadratic-actions}), its source-dependent measure is Gaussian and 
\begin{equation}
W_{\rm IF}[\mathbf J]
=
-i\, \log
\left.
\left\{
\exp\left[
\frac{i}{2}\, 
\frac{\delta}
{\delta\boldsymbol\varphi }\cdot
\mathbb K_{\cal C}^{-1}\cdot
\frac{\delta}{\delta\boldsymbol\varphi}
\right]
e^{i\, S_{\rm IF}[\boldsymbol\varphi]}
\right\}
\right|_{
\boldsymbol\varphi=\mathbf\Phi_{\cal C}[\mathbf J]=-\mathbb K_{\cal C}^{-1}\cdot{\bf J}
}.\label{eq:WIFsk}
\end{equation}
Consequently, all
closed system fluctuation corrections can be evaluated by Wick
contractions, even when $S_{\rm IF}$ is nonquadratic.
\item Conversely, if the closed system is generic but the microscopic
influence action  $S_{\rm IF}$ is quadratic  (\ref{eq:microscopic-quadratic-actions}), then the influence insertion becomes a Gaussian functional-differential
operator:
\begin{equation}
     W_{\rm IF}[\mathbf J] 
    =-i\,\log\left\{ e^{-i\,W_{\cal C}[\mathbf J]}\;
      \displaystyle
      \exp\left[
        -\frac{i}{2}\,
        \frac{\delta}{\delta\mathbf J}
        \cdot \mathbb K_{\rm IF}\cdot
        \frac{\delta}{\delta\mathbf J}
      \right]
      e^{i\,W_{\cal C}[\mathbf J]}
    \right\} .
\label{eq:WIF-quadratic-influence}
\end{equation}
In this case the influence operator is Gaussian, but its action depends
on the complete hierarchy of connected closed system correlation
functions. 
\end{itemize}
When both actions are quadratic, and omitting source independent
normalization terms, the two representations reduce to
\begin{equation}
    W_{\rm IF}[\mathbf J]
    =
    -\frac{1}{2}\, 
    \mathbf J\cdot
    \left[
      (\mathbb K_{\cal C}+\mathbb K_{\rm IF})^{-1}
      -
      \mathbb K_{\cal C}^{-1}
    \right]
    \cdot\mathbf J,
\end{equation}
reproducing
Eq.~\eqref{eq:B-from-microscopic-kernels}.
 
\subsection{Computational perspective}

The additive decomposition of the open system CGF,
Eq.~\eqref{eq:defIF}, allows the closed system and environment induced
contributions to be approximated separately and subsequently combined.
When $W_{\rm IF}$ can be evaluated directly from the microscopic
influence insertion and known closed system correlation functions, as
in Eq.~\eqref{eq:WIFoperator}, $\Gamma_{\rm IF}$ can be constructed
separately and subsequently combined with $\Gamma_{\cal C}$ through
the stationary composition law \eqref{eq:central-composition}. This
avoids performing the Legendre transformation of the complete
open system generating functional. The main cases in which this
strategy may be useful are the following:

\begin{itemize}

\item \emph{Gaussian environments with linear coupling.}
For a Gaussian environment coupled linearly to the system, the
microscopic influence action $S_{\rm IF}$ is quadratic. In this case,
Eq.~\eqref{eq:WIF-quadratic-influence} expresses $W_{\rm IF}$ in terms
of   $W_{\cal C}$ and the known environmental
kernel $\mathbb K_{\rm IF}$ . This setting includes harmonic baths, free thermal fields, and
Gaussian noise.

\item \emph{Weak system environment coupling.}
If $S_{\rm IF}=O(g^2)$, the cumulant expansion (\ref{eq:WIFoperator}) gives
\begin{equation}
W_{\rm IF}[\mathbf J]
=
\left\langle S_{\rm IF}\right\rangle_{{\cal C},\mathbf J}
+
\frac{i}{2}
\left\langle S_{\rm IF}^{\,2}\right\rangle_{{\cal C},\mathbf J}^{\rm conn}
+
\cdots .
\label{eq:WIF-cumulant-expansion}
\end{equation}
For a polynomial influence action, only a finite hierarchy of
closed system correlation functions is required at any fixed order in
$g$.  For example, at leading order a
quadratic influence action depends only on the closed system one  and
two point functions.

\item \emph{Gaussian closed systems.}
If the closed system action ${\cal S}_{\cal C}$ is quadratic, all source dependent
closed system averages can be evaluated by Wick contractions, as in
Eq.~\eqref{eq:WIFsk}. The calculation can remain manageable even when
$S_{\rm IF}$ is nonquadratic, provided that it is polynomial or treated
perturbatively.

\item \emph{Previously determined closed system correlators.}
The method is especially useful when $W_{\cal C}$,
$\Gamma_{\cal C}$, or the relevant closed system correlation functions
are already known, as from perturbation theory,  large-$N$ expansion,
functional renormalization group methods, or numerical calculations.
The same closed system information can then be reused for different
environments without recomputing the complete open system generating
functional.

\item \emph{Semiclassical or large $N$ regimes.}
When closed system fluctuations are suppressed and the
source dependent measure is sharply concentrated around its mean field,
one has, at leading order,
\begin{equation}
W_{\rm IF}[\mathbf J]
\simeq
S_{\rm IF}
\!\left[
\mathbf\Phi_{\cal C}[\mathbf J]
\right]
+
\text{fluctuation corrections}.
\label{eq:WIF-semiclassical}
\end{equation}
The corrections can then be organized systematically in powers of
$\hbar$ or $1/N$. The subsequent Legendre transformation is still
required, so this approximation does not imply
$\Gamma_{\rm IF}=S_{\rm IF}$.

\end{itemize}
In contrast, a computational advantage is not generally expected when
the environment is strongly coupled and non Gaussian, the required
closed system correlation hierarchy is unknown, or long-time secular
effects require substantial resummation. Similarly, if $W_{\rm IF}$ is
obtained only from the difference $W_{\cal O} - W_{\cal C}$ after
$W_{\cal O}$ has already been calculated, stationary composition
remains an exact structural identity but does not by itself simplify
the calculation.

\subsection{Effective potentials}

As a general nonlinear realization of the stationary composition principle, we
consider static homogeneous   SK  field
configurations,
\begin{equation}
\Phi_a(x)=\phi_a,
\qquad
\Psi_a(x)=\psi_a,
\qquad
a=1,2.
\end{equation}
For such configurations, the effective actions reduce to the
corresponding effective potentials,
\begin{equation}
\Gamma_{\cal C}
\longrightarrow
U_{\cal C},
\qquad
\Gamma_{\rm IF}
\longrightarrow
U_{\rm IF},
\end{equation}
where overall spacetime volume factors have been omitted since they play
no role in the stationary condition. 
The auxiliary functional becomes
\begin{equation}
F( \phi,\psi)
=
U_{\cal C}(\psi)
+
U_{\rm IF}(\phi-\psi),
\label{eq:potential_auxiliary}
\end{equation}
and the stationary configuration satisfies
\begin{equation}
\frac{\partial F(\phi,\psi)}
{\partial\psi }
=
0,\quad \to\quad
\frac{\partial U_{\cal C}(\psi)}
{\partial\psi }
=
\left.\frac{\partial U_{\rm IF}(\chi)}
{\partial\chi }\right |_{\chi =(\phi -\psi )}.
\label{eq:potential_force_balance}
\end{equation}
Unlike the general stationary condition, which is a functional
integro-differential equation, Eq.~\eqref{eq:potential_force_balance}
is a {\it nonlinear algebraic system} for the two SK variables $\psi_a$.
Although it need not admit a closed form solution, it has a simple
physical interpretation: the stationary splitting $\psi_\star$ is the
configuration at which the generalized forces generated by the
closed system and influence effective potentials balance one another.
The local existence and uniqueness of the stationary branch are
controlled by the Hessian of the auxiliary functional,
\begin{equation}
H_{ab}(\phi, \psi)
\equiv
\frac{\partial^2F}
{\partial\psi_a\partial\psi_b}
\label{eq:potential_hessian}
\end{equation}
Whenever
\begin{equation}
\det H\neq0,
\end{equation}
the implicit-function theorem guarantees a locally unique stationary
solution,
\begin{equation}
\psi_\star
=
\psi_\star(\phi),
\end{equation}
which determines the composed effective potential,
\begin{equation}
U_{\cal O}(\phi)
=
U_{\cal C}
\!\left(\psi_\star(\phi)\right)
+
U_{\rm IF}
\!\left(\phi-\psi_\star(\phi)\right).
\label{eq:potential_composition}
\end{equation}
This example provides the nonlinear counterpart of the quadratic
stationary composition discussed previously. 
Points at which $H$ becomes non invertible may mark the onset of a
multibranch structure in the auxiliary stationary problem.

   \section{Conclusion}

In this work, we have shown that the additive decomposition of connected
generating functionals,
\[
W_{\cal O}
=
W_{\cal C}
+
W_{\rm IF},
\]
is mapped by the Legendre transformation into a stationary composition
of effective actions,
\[
\Gamma_{\cal O}
=
\Gamma_{\cal C}
\circledast
\Gamma_{\rm IF}.
\]
The construction provides a formal SK analogue of the
infimal convolution structure of convex duality.  Rather than minimizing over an
auxiliary variable, the composition is defined through a stationary
condition that determines a stationary decomposition of the total mean
field, Fig.~\ref{fig:StationaryComposition}. We further showed that this construction is decomposition
covariant: different partitions of the connected dynamics into
closed system and influence contributions lead to the same open system
effective action.
The two realizations considered here illustrate the linear and nonlinear forms of the composition law. For quadratic effective actions, the construction reduces to an operator inversion, whereas general time-independent actions lead to a nonlinear stationary equation governed locally by a field-dependent Hessian.
 Together, these
examples shown that the variational construction extends naturally from
linear response theory to nonlinear effective actions.
The stationary composition may also support a modular approximation
strategy when $W_{\rm IF}$ can be evaluated directly from the
microscopic influence insertion and known closed system correlation
functions. In such cases, $\Gamma_{\cal C}$ and $\Gamma_{\rm IF}$ can
be constructed separately and combined without directly performing the
complete open system Legendre transformation. This advantage is
model dependent and is absent when $W_{\rm IF}$ is available only after
$W_{\cal O}$ has already been calculated.
Several directions naturally emerge from the present work. 
A first extension concerns higher effective actions, most notably the
two-particle irreducible (2PI) effective action \cite{Cornwall:1974vz, Berges:2004yj, Calzetta:2008iqa}, where both mean fields and propagators become independent variational variables. 
A second direction is the analysis of singular Hessians and the associated
multibranch stationary structure, which is expected to play an important
role in phase transitions, metastability, and dissipative critical
phenomena. 
Finally, applying the stationary composition to explicit open
quantum systems and nonequilibrium quantum field theories (in particular in a cosmological setting \cite{Calzetta:1986ey, Boyanovsky:1994me, Weinberg:2005vy, Lau:2024mqm, Pajer:2026fuo, Colas:2025app} or to describe hydrodynamical behaviour of effective field theories \cite{Galley:2012hx, Grozdanov:2013dba, Andersson:2013jga, Crossley:2015evo}) should provide
a useful framework for practical calculations.
  We hope that this perspective will prove useful for the further
development of nonequilibrium quantum field theory and the effective
description of open quantum systems.


\begin{thebibliography}{99} 
 
 \bibitem{Schwinger:1960qe}
J.~S.~Schwinger,
{\it Brownian motion of a quantum oscillator,}
J. Math. Phys. \textbf{2} (1961), 407-432
doi:10.1063/1.1703727

\bibitem{Keldysh:1964ud}
L.~V.~Keldysh,
{\it Diagram Technique for Nonequilibrium Processes,}
Sov. Phys. JETP \textbf{20} (1965), 1018-1026
doi:10.1142/9789811279461{\_}0007

\bibitem{Bakshi:1962dv}
P.~M.~Bakshi and K.~T.~Mahanthappa,
{\it Expectation value formalism in quantum field theory. 1.,}
J. Math. Phys. \textbf{4} (1963), 1-11
doi:10.1063/1.1703883

\bibitem{Chou:1984es}
K.~C.~Chou, Z.~B.~Su, B.~L.~Hao and L.~Yu,
{\it Equilibrium and Nonequilibrium Formalisms Made Unified,}
Phys. Rept. \textbf{118} (1985), 1-131
doi:10.1016/0370-1573(85)90136-X


\bibitem{Jordan:1986ug}
R.~D.~Jordan,
{\it Effective Field Equations for Expectation Values,}
Phys. Rev. D \textbf{33} (1986), 444-454
doi:10.1103/PhysRevD.33.444
 
\bibitem{Kamenev}
A. Kamenev, {\it Field Theory of Non-Equilibrium Systems} (Cambridge University Press, 2011)

\bibitem{Rammer:2007zz}
J.~Rammer,
{\it Quantum field theory of non-equilibrium states,} Cambridge University Press, Cambridge, 2007.
 
 
\bibitem{Calzetta:2008iqa}
E.~A.~Calzetta and B.~L.~B.~Hu,
{\it Nonequilibrium Quantum Field Theory,}
Oxford University Press, 2009,
ISBN 978-1-009-29003-6, 978-1-009-28998-6, 978-1-009-29002-9, 978-0-511-42147-1, 978-0-521-64168-5
doi:10.1017/9781009290036
 

\bibitem{Feynman:1963fq}
R.~P.~Feynman and F.~L.~Vernon, Jr.,
{\it The Theory of a general quantum system interacting with a linear dissipative system,}
Annals Phys. \textbf{24} (1963), 118-173
doi:10.1016/0003-4916(63)90068-X


\bibitem{Caldeira:1982iu}
A.~O.~Caldeira and A.~J.~Leggett,
{\it Path integral approach to quantum Brownian motion,}
Physica A \textbf{121} (1983), 587-616
doi:10.1016/0378-4371(83)90013-4

\bibitem{Paz:1990jg}
J.~P.~Paz,
{\it Anisotropy dissipation in the early universe: Finite temperature effects reexamined,}
Phys. Rev. D \textbf{41} (1990), 1054-1066
doi:10.1103/PhysRevD.41.1054


\bibitem{Hu:1991di}
B.~L.~Hu, J.~P.~Paz and Y.~h.~Zhang,
{\it Quantum Brownian motion in a general environment: 1. Exact master equation with nonlocal dissipation and colored noise,}
Phys. Rev. D \textbf{45} (1992), 2843-2861
doi:10.1103/PhysRevD.45.2843


\bibitem{Breuer:2007juk} H.~P.~Breuer and F.~Petruccione,
{\it The Theory of Open Quantum Systems,}\\
Oxford University Press, 2007,
ISBN 978-0-19-170634-9, \\ 978-0-19-921390-0
doi:10.1093/acprof:oso/9780199213900.001.0001

 
\bibitem{Jackiw:1974cv}
R.~Jackiw,
{\it Functional evaluation of the effective potential,}
Phys. Rev. D \textbf{9} (1974), 1686
doi:10.1103/PhysRevD.9.1686


\bibitem{DeWitt:1964mxt}
B.~S.~DeWitt,
{\it Dynamical theory of groups and fields,}
Conf. Proc. C \textbf{630701} (1964), 585-820
  
\bibitem{Cornwall:1974vz}
J.~M.~Cornwall, R.~Jackiw and E.~Tomboulis,
{\it Effective Action for Composite Operators,}
Phys. Rev. D \textbf{10} (1974), 2428-2445
doi:10.1103/PhysRevD.10.2428

\bibitem{Berges:2004yj}
J.~Berges,
{\it Introduction to nonequilibrium quantum field theory,}
AIP Conf. Proc. \textbf{739} (2004) no.1, 3-62
doi:10.1063/1.1843591
[arXiv:hep-ph/0409233 [hep-ph]].

 \bibitem{Rock}
R.~T.~Rockafellar,  (1970). {\it Convex Analysis} (PMS-28). Princeton University Press.

\bibitem{Stromberg1994}
T.~Str{\"o}mberg, {\it A study of the operation of infimal convolution}, PhD dissertation, Luleå tekniska Universitet, Luleå, 1994

\bibitem{AndersonDuffin1969}
W.~N.~Anderson, Jr. and R.~J.~Duffin,
{\it Series and parallel addition of matrices,}
J. Math. Anal. Appl. \textbf{26} (1969), 576--594,
doi:10.1016/0022-247X(69)90200-5.

\bibitem{Kubo:1957mj}
R.~Kubo,
{\it Statistical mechanical theory of irreversible processes. 1. General theory and simple applications in magnetic and conduction problems,}
J. Phys. Soc. Jap. \textbf{12} (1957), 570-586
doi:10.1143/JPSJ.12.570

 \bibitem{Kleinert:2004ev}
H.~Kleinert,
{\it Path Integrals in Quantum Mechanics, Statistics, Polymer Physics, and Financial Markets,}(World Scientific, 2009).
 


\bibitem{Weinberg:2005vy}
S.~Weinberg,
{\it Quantum contributions to cosmological correlations,}
Phys. Rev. D \textbf{72} (2005), 043514
doi:10.1103/PhysRevD.72.043514
[arXiv:hep-th/0506236 [hep-th]].


\bibitem{Su:1987pi}
Z.~b.~Su, L.~y.~Chen, X.~t.~Yu and K.~c.~Chou,
{\it Influence functional and closed-time-path Green's function,}
Phys. Rev. B \textbf{37} (1988), 9810-9812
doi:10.1103/PhysRevB.37.9810
 
 
 \bibitem{Boyanovsky:1994me}
D.~Boyanovsky, H.~J.~de Vega, R.~Holman, D.~S.~Lee and A.~Singh,
{\it Dissipation via particle production in scalar field theories,}
Phys. Rev. D \textbf{51} (1995), 4419-4444
doi:10.1103/PhysRevD.51.4419
[arXiv:hep-ph/9408214 [hep-ph]].

\bibitem{BenTov:2021jsf} Y.~BenTov, 
{\it SK 
path integral for the quantum harmonic oscillator}, 
[arXiv:2102.05029 [hep-th]].

\bibitem{Proko:2017}
D. Glavan and T. Prokopec,   {\it A pedestrian introduction to nonequilibrium QFT}, lecture notes, Utrecht University (2017),\\ https://webspace.science.uu.nl/~proko101/LecturenotesNonEquilQFT.pdf
 

\bibitem{Melo:2021mbd}
J.~F.~Melo,
{\it The propagator matrix reloaded,}
SciPost Phys. Core \textbf{6} (2023), 019\\
doi:10.21468/SciPostPhysCore.6.1.019
[arXiv:2112.09119 [hep-th]].

\bibitem{Barvinsky:2023jkl}
A.~O.~Barvinsky and N.~Kolganov,
{\it Nonequilibrium Schwinger-Keldysh formalism for  \it density matrix states: Analytic properties and implications in cosmology,}
Phys. Rev. D \textbf{109} (2024) no.2, 025004
doi:10.1103/PhysRevD.109.025004
[arXiv:2309.03687 [hep-th]]

\bibitem{Launay:2024trh}
Y.~L.~Launay, G.~I.~Rigopoulos and E.~P.~S.~Shellard,
{\it Quantitative classicality in cosmological interactions during inflation,}
JCAP \textbf{05} (2025), 071
doi:10.1088/1475-7516/2025/05/071
[arXiv:2412.16143 [gr-qc]].

\bibitem{Calzetta:1986ey}
E.~Calzetta and B.~L.~Hu,
{\it Closed Time Path Functional Formalism in Curved Space-Time: Application to Cosmological Back Reaction Problems,}
Phys. Rev. D \textbf{35} (1987), 495
doi:10.1103/PhysRevD.35.495


\bibitem{Lau:2024mqm}
P.~H.~C.~Lau, K.~Nishii and T.~Noumi,
{\it Gravitational EFT for dissipative open systems,}
JHEP \textbf{02} (2025), 155
doi:10.1007/JHEP02(2025)155
[arXiv:2412.21136 [hep-th]].

\bibitem{Pajer:2026fuo}
E.~Pajer,
{\it Lectures on Open Systems and Cosmology:~Master Equation, Schwinger{\textendash}Keldysh and Open Effective Field Theories,}
[arXiv:2607.14351 [hep-th]].

\bibitem{Colas:2025app}
T.~Colas,
{\it Lectures on Open Effective Field Theories,}
[arXiv:2510.00140 [hep-th]].

  \bibitem{Grozdanov:2013dba}
S.~Grozdanov and J.~Polonyi,
{\it Viscosity and dissipative hydrodynamics from effective field theory,}
Phys. Rev. D \textbf{91} (2015) no.10, 105031
doi:10.1103/PhysRevD.91.105031
[arXiv:1305.3670 [hep-th]].

\bibitem{Andersson:2013jga}
N.~Andersson and G.~L.~Comer,
{\it A covariant action principle for dissipative fluid dynamics: From formalism to fundamental physics,}
Class. Quant. Grav. \textbf{32} (2015) no.7, 075008
doi:10.1088/0264-9381/32/7/075008
[arXiv:1306.3345 [gr-qc]].

\bibitem{Crossley:2015evo}
M.~Crossley, P.~Glorioso and H.~Liu,
{\it Effective field theory of dissipative fluids,}
JHEP \textbf{09} (2017), 095,
doi:10.1007/JHEP09(2017)095,
[arXiv:1511.03646 [hep-th]].

\bibitem{Galley:2012hx}
C.~R.~Galley,
{\it Classical Mechanics of Nonconservative Systems,}
Phys. Rev. Lett. \textbf{110} (2013) no.17, 174301
doi:10.1103/PhysRevLett.110.174301
[arXiv:1210.2745 [gr-qc]].
  
\end{thebibliography}
    \end{document}